\documentclass[a4paper,11pt]{article}
\pdfoutput=1 

\usepackage{jheppub} 

\usepackage[T1]{fontenc} 

\usepackage{physics}
\usepackage{amsmath}
\usepackage{dsfont}
\usepackage{graphicx}
\usepackage{bbold}
\graphicspath{ {./1 /} }

\usepackage{physics}
\usepackage[dvipsnames, svgnames,  x11names ]{xcolor}
\usepackage{tikz}
\usetikzlibrary{arrows,snakes,backgrounds}
\newcommand{\RNum}[1]{\uppercase\expandafter{\romannumeral #1\relax}}

\usepackage{physics}
\usepackage[dvipsnames, svgnames,  x11names ]{xcolor}
\usepackage{tikz}

\title{\boldmath
A CFT dual for evaporating black holes: boundary continuous matrix product states
}

\author[]{Niloofar Vardian}

\affiliation{SISSA, International School for Advanced Studies, via Bonomea 265, 34136 Trieste, Italy}

\emailAdd{nvardian@sissa.it}
\emailAdd{niloofarvardian@gmail.com}

\abstract{
Tensor network states, especially Matrix Product States (MPS), are crucial tools for studying how particles in large quantum systems are entangled with each other. MPS are particularly effective for modeling systems in one-dimensional space. Their continuous version, known as continuous Matrix Product States (cMPS), extends this approach to more complex quantum field theories that describe systems with an infinite number of interacting particles.
This paper introduces a novel extension, boundary continuous Matrix Product States (BCMPS), which incorporate boundary states from conformal field theory (CFT). We construct BCMPS and explore their potential holographic duals, linking them to black hole microstates with end-of-the-world branes in AdS/CFT. This connection hints at a deeper relationship between tensor networks and spacetime geometry, potentially offering new insights into the interplay between quantum information and gravity.
}

\begin{document}
\maketitle
\flushbottom

\section{Introduction}

In recent years, tensor network states have arisen as the entanglement-based ansatz. They are based on renormalization group ideas and later on developed using the concepts from quantum information theory.

The most important class of tensor networks is Matrix Product State (MPS). By construction, the class of MPS obeys the entropy/area law. The class of MPS provides an efficient class of variational ansatz to approximate the ground state of the local Hamiltonian. Understanding the low-energy behavior of many-body quantum systems is one of the major challenges of modern physics. Thus, the class of MPS is a very useful technique in both high-energy and condensed-matter physics. The MPS is just efficient in 1 spatial dimension. It is a consequence of the area law. There is a generalization of the class of MPS to Projected Entangled Pair State (PEPS) in higher dimensions.

The MPS can be generalized to study non-relativistic quantum field theory (QFT) in one spatial dimension. Thus, quantum fields, with infinite degrees of freedom, in principle can be described by defining continuous MPS (cMPS) \cite{verstraete2010continuous}. 
It is proven in \cite{verstraete2010continuous} that cMPS can be understood as the continuous limit of MPS.
The class of cMPS can be described by the set of matrices as
\begin{equation}
  \ket{ \psi [ Q, R]} = Tr_{aux} \big\{B \mathcal{P} \exp \int _{-L/2} ^ {L/2}  dx 
\big( Q(x) \otimes I +R(x) \otimes \psi ^\dagger (x)\big) \big\} \ket{\Omega} 
\end{equation}
while $ R(x) $ and $ Q(x) $ are finite dimensional $D \times D$ matrices that $D$ is the bond dimension of the state. Moreover, we have $\ket{\Omega}$ that is the vacuum of the nonrelativistic theory which has no spatial entanglement and 
\begin{equation}
 \psi (x) \ket{\Omega} =0 \qquad \qquad \forall x \in [0, L]   
\end{equation}
while $\psi (x)$ is localized field operators of the original non-relativistic theory.

It is possible to build these sets of QFT states in the lab \cite{barrett2013simulating}.  In order to do it, one needs to use the interpretation of the cMPS class as the action of one unitary in the cavity-QED theory. The cMPS is a variational method exploiting the natural physics of cavity-QED.
The cMPS can actually be constructed under a sequential preparation prescription in the same way as the MPS thus appearing in a natural way as proved in \cite{osborne2010holographic}.
Using the theory of continuous measurement, in which a quantum system is subjected to a sequence of weak measurements of POVM. The class of cMPS can be rewritten as 
\begin{equation}\label{1111}
 \ket{ \psi [ Q, R]}=  \mathcal{P} \exp -i \int_0^L ds \Big( K(s) \otimes I + iR(s) \otimes O^\dagger(s) -i R^\dagger (s) \otimes O(s)\Big) \ket{\Omega}
\end{equation}
while 
\begin{equation}
Q= - \frac{1}{2} R^\dagger R-i K.
\end{equation}
In the theory of continuous measurement, one can interpret the parameter $x$ in \eqref{1111} as time. 
The QFT system and ancilla together evolved under the Hamiltonian 
\begin{equation}
H(t) = K(t) \otimes I + i R(t) \otimes \psi ^\dagger (t) - i R^\dagger (t) \otimes \psi (t) 
\end{equation}
$K(t)$ is the Hamiltonian of the ancilla while the interacting Hamiltonian is 
\begin{equation}
H_{int} (t) =  i R(t) \otimes \psi ^\dagger (t) - i R^\dagger (t) \otimes \psi (t) .
\end{equation}
The exchange of the particle between the ancilla and QFT  is controlled by the matrices $R$. 

Recently, the class of cMPS has been generalized to a new class of tensor networks as relativistic cMPS (RCMPS) which is suitable for the relativistic QFT \cite{tilloy2021relativistic, Vardian:2022vwf}. 
Knowing this, we are interested in building some other new class of cMPS.

In $ 1+1$ CFT, we have the class of boundary states denoted as $ \ket{B_s}$. The regularized one can be built by Euclidean time evolution of the state, we denote it as 
\begin{equation}
\ket {B_{s, \beta}} \propto  e^{- \beta H} \ket{B_s}.
\end{equation}
In \cite{miyaji2015boundary}, it has been shown that the regularized boundary states have no spatial entanglement. Therefore, one can think about this class of states as an alternative to the non-relativistic vacuum, i.e. $ \ket {\Omega}$. If one finds a set of operators that kill the $ \ket{B_{s, \beta}} $, we can repeat the same logic of the construction of cMPS to build a new class of cMPS. 
Later, we discuss that this set of operators exists 
\begin{equation}
T_{tx} \ket{B_{s, \beta}}=0
\end{equation}
while $T$ is the energy-momentum tensor. Thus, the new class of cMPS can be built as 
\begin{equation}
 \ket{ \psi_B [ Q, R]} = Tr_{aux} \big\{B \mathcal{P} \exp \int _{-L/2} ^ {L/2}  dx 
\big( Q(x) \otimes I +R(x) \otimes T_{tx} ^\dagger (x)\big) \big\} \ket{B_{s, \beta}}
\end{equation}
We call this new class of tensor network boundary cMPS (BCMPS).

In modern physics, one of the most fascinating connections is between entanglement and spacetime geometry.  
Evidence for this relationship first appeared in the Bekenstein-Hawking formula \cite{bekenstein1973black, hawking1975particle} relating the entropy of the black hole to the surface area of the horizon AdS/CFT provides a concrete realization of this idea while entanglement has come to play a fundamental role in attempts to reconstruct the bulk from CFT data. 

From a very different perspective, tensor networks are entanglement-based ansatz. The question that has arisen now is if some class of tensor network can present some geometries. 
In particular, for one class of tensor network called Multi-scale Entanglement Renormalization Ansatz (MERA) it has been shown. The class of MERA is related to the AdS geometry and in high-energy physics, people work on AdS/MERA connection \cite{bao2015consistency}. 

Now, for our new class i.e. BCMPS: can we find a holographic dual geometry?

It has been shown in \cite{takayanagi2011holographic,cooper2019black} that regularized boundary states for some suitable range of parameters are dual to the black hole microstates with an end of the word (EOW) branes.
Thus, in AdS$_3$/ CFT$_2$, the states of the BCMPS in the dual CFT correspond with a proper geometry of black hole microstates which ends on the EOW branes which is coupled with an ancilla. The ancilla must be the system that absorbs the Hawking radiation. 
From the cavity-QED interpretation of the class of cMPS, one can find the interaction between ancilla and dual theory on the boundary. This term shows the physics we need to describe the absorption of Hawking radiation in the ancilla.

The structure of the paper is as follow, in Chapters 2 and 3,  we review the class of cMPS and the regularized boundary states. In Chapter 4, we build a class of BCMPS  and in the end in Chapter 5, we discuss the holographic dual of the class of BCMPS.

\section{Continuous Matrix Product States}

The family of MPS \cite{fannes1992finitely, klumper1991equivalence, klumper1993matrix} is probably the most famous example of Tensor Network states. This is because it is behind some very powerful methods to simulate the one-dimensional quantum many-body systems.

\begin{figure}[h]
    \centering
    \includegraphics[width=.8\textwidth]{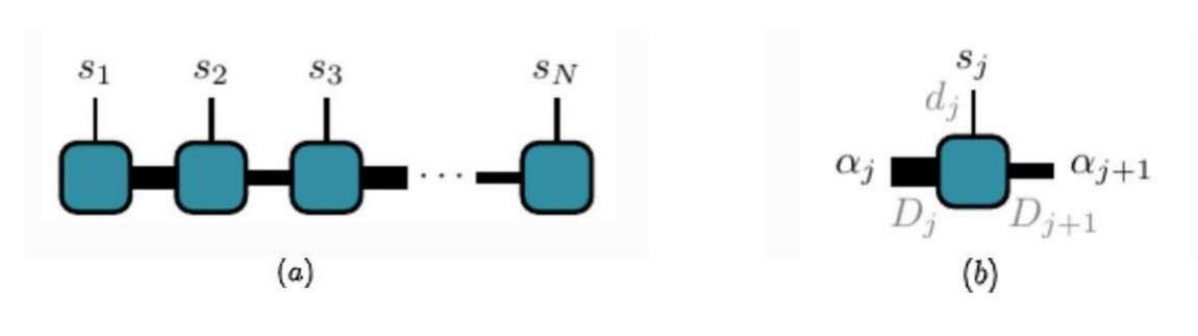}
    \caption{ (a) MPS class of Tensor Network. (b) A rank-3 tensor. }
    \label{fig1}
\end{figure}

MPS are a special class of tensors that can be written as products over many rank-3 tensors, See Fig. \ref{fig1}. Each square have represent a rank-3 tensor (rank-2 for the left and right boundaries) $ A^{s_j}_{\alpha_j, \alpha_{j+1}}$.
The vertical lines represent the physical indices and the horizontal lines are called ancillary indices.
The MPS diagram in Fig. \ref{fig1}
is a rigorous representation of the mathematical expression
\begin{equation}
    C^{s_1s_2...s_N} = \sum _{\{\alpha\}} A _{\alpha_1}^{s_1}[1] A _{\alpha_1 \alpha_2}^{s_2}[2] ... A _{\alpha_{N-2} \alpha_{N-1}}^{s_{N-1}}[N-1] A_{\alpha_{N-1}}^{s_{N}}[N].
\end{equation}
where $ \alpha_i \in \{1,..., D_i\}$.
MPS can represent any quantum state of the many-body Hilbert space just by increasing sufficiently the value of $D_i$.
To see that, consider a quantum many-body system of $N$ particles. Let us take
\begin{equation}
    \ket{\psi} = \sum _{s_1,...,s_N=0} ^{d-1} C^{s_1,...,s_N} \ket{s_1} \otimes ... \otimes \ket{s_N}
\end{equation}
be the state of the $N$ qudit (d-dimensional quantum systems). The state is completely specified by knowledge of the rank-N tensor $C$. One can obtain the MPS representation by breaking the wave function into small pieces.  By starting from 
the first index and split it out from the rest and perform a singular value decomposition, we can get the Schmidt decomposition.
We can now perform successive singular value decomposition along the indices and obtain that 
\begin{equation}
    \ket{\psi} = \sum _{s_1,...,s_N=0} ^{d-1}  A^{s_1}[1] A^{s_2}[2] ... A^{s_{N-1}}[N-1] A^{s_{N}}[N] \ket{s_1} \otimes ... \otimes \ket{s_N}.
\end{equation}
We can redefine the first and last tensors as 
\begin{equation}
    \begin{split}
      A^{s_1}[1] &\longrightarrow \bra{v_L} A^{s_1}[1]
      \\
       A^{s_{N}}[N] &\longrightarrow A^{s_{N}}[N] \ket{v_R}.
    \end{split}
\end{equation}
Thus, the tensors $ A^{s_{1}}[1]$ and $A^{s_{N}}[N]$ are the rank-2 tensors as well. Therefore, we have
\begin{equation}
    C^{s_1,...,s_N} = \bra{ v_L} A^{s_1}[1] A^{s_2}[2] ... A^{s_{N-1}}[N-1] A^{s_{N}}[N] \ket{v_R}.
\end{equation}
Moreover, one can take the periodic boundary condition by putting $ N+1 \equiv 1$. As a result
\begin{equation}
    C^{s_1,...,s_N} = Tr\big[ A^{s_1}[1] A^{s_2}[2] ... A^{s_{N-1}}[N-1] A^{s_{N}}[N] \big].
\end{equation}
For states that are translationally symmetric, we can choose
\begin{equation}
   A^s[1] = A^s[2]=...= A^s[N] \equiv A^s 
\end{equation}
and take all $D_i$ equal to single $D$. In the end, the MPS representation can be obtained as 
\begin{equation}
    \ket{\psi} = \sum _{s_1,...,s_N=0} ^{d-1} Tr\big[B A^{s_1} A^{s_2}... A^{s_{N-1}} A^{s_{N}}\big] \ket{s_1} \otimes ... \otimes \ket{s_N}
\end{equation}
where the information about the boundary conditions is encoded in the matrix $B$. We have $ B=I$ in the case of periodic boundary conditions and $ B = \ket{v_R} \bra{v_L}$ in the case of open boundary conditions.
As it is mentioned, every state can be generally written in the MPS form with $D$ growing exponentially with the particle number $N$. However, MPS is practical when $D$ is small. It is particularly useful for dealing with the ground state of a one-dimensional quantum spin model.

The continuum limit of MPS known as countinuous MPS (cMPS) was proposed in \cite{verstraete2010continuous} by Verstraete and Cirac.
It is originally introduced as a variational ansatz for the ground state of non-relativistic QFT Hamiltonians in $1+1$ dimensions.

To find a
generalization of MPS in the continuum limit, one can approximate the QFT on a line of length $L$ by a lattice with lattice spacing $ \epsilon$ and $ N = L/\epsilon$ sites.
At each site of the lattice, there is a bosonic (or fermionic) mode $ a_i$ obeys the commutation relation $ [a_i , a_j^\dagger]_\pm = \delta_{ij}$. Therefore, the Hilbert space spanned by $ \{ \ket{n_i}\}$ while $ \ket{n_i}$ corresponding to having $ n_i$ particles on that site.
For the many-body state we have
\begin{equation}
 \ket{ i_1 ,i_2, ... , i_N} =  a_1 ^{\dagger i_1}  a_2 ^{\dagger i_2} ... a_N ^{\dagger i_N} \ket{\boldsymbol{0}},  
\end{equation}
where $ \ket{\boldsymbol{0}} = \otimes _ {n =1} ^N \ket{0}_n $
is the vacuum that 
\begin{equation}
    a_j \ket{\boldsymbol{0}} =0 \qquad \forall j.
    \end{equation}
On the lattice, we can define a certain family of MPS as 
\begin{equation}
    \begin{split}
         A^0  _i &= I + \epsilon Q( i\epsilon)
         \\
         A^n _i &= \frac{1}{n!} \big( \sqrt{\epsilon} R( i \epsilon)\big)^n\qquad n \geq 1.
    \end{split}
\end{equation}
For higher $n$, the matrices $A^n$
have been determined by the requirement that  a
doubly occupied site gives the same physics as 2 bosons 
on 2 neighboring sites in the limit $ \epsilon \rightarrow 0$.
By taking the $ \epsilon \rightarrow 0$ limit of this specific class of MPS, we can find the class of cMPS as

\begin{equation}
 \ket{ \psi [ Q, R]} = Tr_{aux} \big\{B \mathcal{P} \exp \int _{-L/2} ^ {L/2}  dx 
\big( Q(x) \otimes I +R(x) \otimes \psi ^\dagger (x)\big) \big\} \ket{\Omega}      
\end{equation}
where $ Tr_{aux}$ denotes a partial trace over the auxiliary system where the matrices $ Q$ and $R$ act. For the translational invariant cMPS the matrices $Q,~ R$ are position independent. 
The field $ \psi(x)$ is the continuum limit of the rescale modes 
$ \psi (i \epsilon) = a_i / \sqrt{\epsilon}$, that
satisfying $ [\psi (x), \psi ^ \dagger (y)]_\pm = \delta (x-y)$, and $ \ket{\Omega}$, the empty vacuum 
is the continuum limit of $ \ket{\boldsymbol{0}}$ that 
defined as 
\begin{equation}
     \psi(x) \ket{\Omega} =0 \qquad\qquad \forall x.
\end{equation}

One can express the expectation value of local operators and in particular, the Hamiltonian on the cMPS representation of the ground state in terms of the matrices $Q$ and $R$.
Specifically, all normal ordered correlation functions of local field operators can be deduced from a generating functional as
\begin{equation}
    \langle : F[ \psi^\dagger (x), \psi (y)]: \rangle = F\big[ \frac{\delta}{\delta \bar{j} (x)}, \frac{\delta}{\delta j (y)} \big] \mathcal{Z}_{ \bar{j}, j} \big|_{\bar{j}, j =0}, 
\end{equation}
while its explicit form can be given in terms of the cMPS matrices $Q$ and $R$ as 
\begin{equation}\label{7}
    \mathcal{Z}_{ \bar{j}, j} = Tr \Big\{ B  \otimes \Bar{B}\mathcal{P} \exp \big[\int dx~  T+ j(x)~ R\otimes I + \bar{j}(x)~ I \otimes \bar{R}\big] \Big\}
\end{equation}
where
\begin{equation}
  T = Q \otimes I + I \otimes \bar{Q} + R \otimes \bar{R}  
\end{equation}
is the cMPS transfer matrix \cite{haegeman2013calculus}.

In order to find the cMPS approximation of the ground state, it is just needed to minimize the expectation value of the Hamiltonian over the cMPS matrices $Q$ and $R$. After that, correlation functions can be straightforwardly computed.
The cMPS representation has gauge freedom 
\begin{equation}
   \begin{split}
       Q(x) & \longrightarrow g(x) Q(x) g^{-1}(x)- \frac{dg(x)}{dx}g^{-1}(x)
       \\
       R(x) & \longrightarrow g(x) R(x) g^{-1}(x)
   \end{split} 
\end{equation}
that one can use to impose certain conditions on the cMPS matrices, including symmetry conditions. Moreover, for the continuum version, the left orthogonality condition of MPS can be read as
\begin{equation}
  Q(x) + Q^\dagger (x) + R^\dagger (x) R(x) =0  
\end{equation}
for all $x$.
A better approximation of the ground state can be found by increasing $ D$.
In the last decade, several optimization algorithms have been developed to study a number of theories,  both bosonic and fermionic 
\cite{haegeman2010applying, draxler2013particles, quijandria2014continuous, chung2015matrix, quijandria2015continuous, haegeman2017quantum, rincon2015lieb, chung2017multiple, draxler2017continuous, ganahl2017continuous, ganahl2017continuous2, ganahl2018continuous, tuybens2022variational2}.
The cMPS provides an efficient variational ansatz for non-relativistic QFTs. It is not adapted to relativistic theories because of a lack of sensitivity to short-distance behavior.

\section{Regularized Boundary States}

 Regularized boundary states play a crucial role in understanding the behavior of CFTs in the presence of boundaries. These boundary states capture the impact of boundary conditions on the CFT living on the boundary, allowing us to study various physical phenomena related to open quantum systems or the presence of interfaces.

The regularization process involves evolving the boundary state along Euclidean time, effectively smearing out high-energy contributions, and ensuring that the state possesses finite energy. This regularization is essential to render the theory well-defined and to make physical predictions that are consistent and meaningful.

Regularized boundary states are valuable tools in exploring the physics of CFTs in various contexts, such as boundary critical phenomena, quantum entanglement at boundaries, and interface dynamics in condensed matter systems. They provide a natural framework to understand the interplay between bulk and boundary degrees of freedom and the emergence of universal features near the boundary.

Moreover, regularized boundary states facilitate the study of entanglement entropy and entanglement spectra at the interface between different phases of matter, helping to reveal the underlying quantum phase transitions and topological properties of the system. They also find applications in holography, where they correspond to boundary states of the corresponding AdS in the AdS/CFT correspondence, connecting insights from gravity and quantum field theory.

\subsection{Boundary states in 2D CFT}

In a 2D CFT, boundary states are required to fulfill the condition stated in \cite{Cardy:2004hm}
\begin{equation}\label{3}
(L_n - \tilde{L}_n) \ket{B} = 0.
\end{equation}
Here, $L_n$ and $\tilde{L}_n$ represent the Virasoro generators associated with the left and right-moving sectors, respectively, and $\ket{B}$ denotes the boundary state.
Within any Verma module, a straightforward solution to these conditions can be found as follows
\begin{equation}
\ket{I_h} = \sum_{\Vec{k}} |\Vec{k},h\rangle_L \otimes |\Vec{k},h\rangle_R ,,
\end{equation}
Here, $|\Vec{k},h\rangle_L$ is a linear combination of Virasoro descendants of the primary state $|h\rangle$, which is characterized by an infinite-dimensional vector $\Vec{k} = (k_1, k_2, ...)$ with non-negative integer components.
We recognize these states by considering descendants of the following form:
\begin{equation}\label{2d-basis}
... L_{-n}^{K_{n}} ... L_{-1}^{K_{1}} \ket{h}_L,
\end{equation}
where we construct an orthonormal basis, ensuring that $_{L}\langle\Vec{k},h|\Vec{k'},h\rangle_L = \delta_{\Vec{k}, \Vec{k'}}$.

The state $ \ket{I_h}$ is referred to as the Ishibashi state associated with the primary state $ \ket{h}_L$, where the states $|\Vec{k},h\rangle$ represent descendants built upon the primary state labeled by $h$. It is readily apparent that the following relation holds:
\begin{equation}
L_n |I_h\rangle = \tilde{L}_n |I_h \rangle .
\end{equation}
The Ishibashi states exhibit maximum entanglement between the left-moving and right-moving sectors. Furthermore, linear combinations of Ishibashi states also satisfy the constraint \eqref{3}.

Physical boundary states are expressed as specific linear combinations of Ishibashi states, referred to as Cardy states:
\begin{equation}
\ket{B_a} = \sum _h C_{a,h} \ket{I_h} ,.
\end{equation}
To be considered as physically valid, these boundary states must fulfill a consistency condition related to open-closed duality, which emerges from the partition function on a finite cylinder, as described in \cite{Cardy:2004hm}.

The Cardy states become singular due to the divergent norm of the Ishibashi states. To address this, one can introduce regularized boundary states by evolving them in Euclidean time:
\begin{equation}\label{2}
\ket{B_{a,\beta}} = e^{- \frac{\beta}{4} H_c} \ket{B_a},
\end{equation}
where $ \beta$ is a positive constant and $H_c = L_0 + \tilde{L}_0 - \frac{c}{12}$. This regularization ensures that the state \eqref{2} remains space-translationally invariant on the circle but becomes time-dependent.

\subsection{Entanglement entropy of boundary states}\label{3.2}

Let us start with the massless Dirac fermion theory in 2d. The system is in the boundary state $e^{-\epsilon H}\ket{B}$ in either
the Dirichlet or Neumann boundary condition. We need to calculate the entanglement entropy $S_A$ when $A$ is an interval.
In \cite{miyaji2015boundary} based on the work in \cite{takayanagi2010measuring}, the entanglement entropy $S_A$ for the corresponding set of states has been calculated. It has been shown that 
\begin{equation}
    S_A \sim O(1).
\end{equation}
Since we need to introduce the cut-off $\epsilon$ used as the damping factor, we have the ambiguity of shifting $\epsilon$. This means that $O(1)$ entropy can be changed by the choice of the UV cut-off and  this is enough to
argue that boundary states essentially have no real-space entanglement.

\subsection{Holographic boundary states}\label{3.3}
In holographic theories, the gravity path integral can be related to the CFT path integral through the AdS/CFT correspondence. Consequently, by selecting an appropriate state with a well-understood gravity prescription for handling the boundary condition at the initial Euclidean time, we can derive the corresponding geometries. Cooper et al. \cite{cooper2019black} discussed the method of describing boundary states by initiating with the Thermofield Double (TFD) state of two CFTs denoted as L and R
\begin{equation}
    \ket{{\rm TFD}(\beta /2)} = \frac{1}{Z} \sum_i e^{-\beta E_i /4} \ket{E_i}_L \otimes \ket{E_i}_R.
\end{equation}
Subsequently, we perform a projection of the TFD state onto a specific pure state $\ket{B}$ belonging to the left CFT.
Hence, the outcome is a pure state in the right Conformal Field Theory (CFT) represented as
\begin{equation}
\ket{B_{a,\beta}} = \frac{1}{Z} \sum_i e^{-\beta E_i /4} \langle B_a \ket{E_i} \ket{E_i}.
\end{equation}

In case of a sufficiently high temperature, the TFD state corresponds to the maximally extended AdS-Schwarzschild black hole in the bulk according to the duality. The geometry associated with these regularized boundary states is anticipated to encompass a substantial portion of the left asymptotic region.
Thus, the state 
$ \rho \propto  e^{-\epsilon H} \ket{B} \bra{B} e^{-\epsilon H}$, which corresponds to a 2d CFT on a strip of width $ 2 \epsilon$, has a gravity dual described by a section of the Euclidean BTZ black hole. The metric of the Euclidean black hole is given as
\begin{equation}
    ds^2 = R^2 \Big( \frac{h(z) dt^2}{z^2} + \frac{dz^2}{h(z) z^2} + \frac{dx^2}{z^2}\Big), \qquad\qquad h(z)= 1- \frac{\pi^2 z^2}{4 \epsilon^2}
\end{equation}
where $R$ is the AdS radius and the ranges of the coordinates $ (t,z,x)$ are $ -2\epsilon \leq t \leq 2 \epsilon$ with periodicity of $ 4 \epsilon$, $ 0< z \leq 2 \epsilon / \pi $ and $ -\infty < x < \infty$. 
Consequently, in the context of a holographic CFT, this set of regularized boundary states can be considered as microstates of a single-sided black hole.
These black hole microstates can be conceptualized as black holes accompanied by end-of-the-world (EOW) branes positioned on the left side.
Typically, the EOW brane setup manifests as a time-dependent configuration on a macroscopic scale.

\section{Bulding a class of cMPS over the boundary states}

As it has been discussed in Sec. \ref{3.2}, the entanglement entropy of spatial regions in boundary states vanishes. Therefore, there exists a set of operators denoted as $ O(x)$ such that 
\begin{equation}\label{2222}
    O(x) \ket{B_{a,\beta}} =0 \qquad \forall x.
\end{equation}

Let us consider a boundary state $ \ket{B_{a,\beta}}$ in 2 dimensions. One can approximate the QFT on a special line in 2 dimensions by a lattice spacing $ \epsilon$ and $ N = L/ \epsilon$  sites. One set of basis can be given as 
\begin{equation}
    O^{\dagger i_1} ( \epsilon)  O^{\dagger i_2} ( 2\epsilon)...  O^{\dagger i_N} ( N\epsilon)  \ket{B_{a,\beta}}
\end{equation}
and the MPS representation of one such class of states is given as 
\begin{equation}
    \ket{\Psi} = \sum _{i_1, ..., i_N =0}^\infty \Tr \big[ B A^ {i_1} A^{i_2}...A^{i_N}\big]
    O^{\dagger i_1} ( \epsilon)  O^{\dagger i_2} ( 2\epsilon)...  O^{\dagger i_N} ( N\epsilon)  \ket{B_{a,\beta}}.
\end{equation}
Define a specific class of MPS as 
\begin{equation}
    \begin{split}
        A^0& (n\epsilon) = I + \epsilon Q(n \epsilon)
        \\
        A^1& (n \epsilon) = \sqrt{\epsilon} R (n\epsilon) 
        \\
        A^k & (n\epsilon) = A^1(n\epsilon)^k/k!.
    \end{split}
\end{equation}
To find an explicit form of the MPS representation in the continuum limit one can introduce 
\begin{equation}
    \ket{\Psi} = \sum _{n=0}^\infty \ket{\Psi_n}
\end{equation}
while 
\begin{equation}
        \ket{\Psi_n}=   \sum _{i_1+ ...+ i_N =n}\Tr \big[ B A^ {i_1} A^{i_2}...A^{i_N}\big]
    O^{\dagger i_1} ( \epsilon)  O^{\dagger i_2} ( 2\epsilon)...  O^{\dagger i_N} ( N\epsilon)  \ket{B_{a,\beta}}.
\end{equation}
For $ n=0$ and small value of $ \epsilon$ we have 
\begin{equation}
    \begin{split}
        \ket{\Psi_0}& =\Tr \big[B A^0(\epsilon)...A^0(N\epsilon)\big] \ket{B_{a,\beta}} 
        \\
       & = \Tr \big[B(I+ \epsilon Q(\epsilon))...(I+ \epsilon Q(N\epsilon))\big] \ket{B_{a,\beta}}
       \\
       & = \Tr \big[ B\mathcal{P} \exp \big(\sum_{m=1}^N \epsilon Q (m\epsilon)\big)\big] \ket{B_{a,\beta}}
    \end{split}
\end{equation}
where $\mathcal{P} $ is the path order. In the $ \epsilon \rightarrow 0 $ limit, we reach
\begin{equation}
     \ket{\Psi_0} = \Tr \big[ B \mathcal{P} \exp \big(\int _0^L dx Q(x)\big)\big]\ket{B_{a,\beta}}.
\end{equation}
Then we consider $ n=1$ term
\begin{equation}
    \begin{split}
        \ket{\Psi_1} =& \sum_{j=1}^\infty \Tr\big[B A^0(\epsilon) ... A^0((j-1)\epsilon)
     A^1(j\epsilon)
        A^0((j+1)\epsilon)...A^0(N\epsilon)\big] O^\dagger (j\epsilon) \ket{B_{a,\beta}}
        \\
        =& \sum_{j=1}^N \epsilon \Tr \big[ B\mathcal{P}e^{\sum_{m=1}^{j-1} \epsilon Q(m\epsilon)} R(j\epsilon)
        \mathcal{P}e^{\sum_{m=j+1}^{N} \epsilon Q(m\epsilon)} \big]
     O^\dagger (j\epsilon) \ket{B_{a,\beta}}
    \end{split}
\end{equation}
and in the $\epsilon \rightarrow 0$ limit, we get
\begin{equation}
    \ket{\Psi_1} =\int _0 ^L dx~ \Tr \Big[B \mathcal{P } \big\{e^{\int_0^L ds Q(s)} R(x)\big\}\Big] O^\dagger(x) \ket{B_{a,\beta}}.
\end{equation}
For a generic $n$, one can find that 
\begin{equation}
     \ket{\Psi_n} =\frac{1}{n!}\int _0 ^L dx_1 dx_2...dx_n~ \Tr \Big[B \mathcal{P } \big\{e^{\int_0^L ds Q(s)} R(x_1)...R(x_n)\big\}\Big] O^\dagger(x_1)...O^\dagger(x_n) \ket{B_{a,\beta}}.
\end{equation}
Therefore, we find 
\begin{equation}\label{mps}
    \ket{\Psi}= \sum_{n=0}^\infty \int_{0\leq x_1\leq ...\leq x_n\leq L} dx_1...dx_n~ 
      \Phi_n(x_1,...,x_n) O^\dagger(x_1)...O^\dagger(x_n) \ket{B_{a,\beta}}
\end{equation}
while 
\begin{equation}
    \Phi_n(x_1,...,x_n) = \Tr \Big[ B\mathcal{P}\big\{ e^{\int_0^L Q(s) ds}R(x_1)...R(x_n)\big\}\Big].
\end{equation}
We can rewrite \eqref{mps} as
\begin{equation}
    \begin{split}
        \ket{\Psi}= & \sum_{n=0}^\infty \frac{1}{n!} \int _0^L dx_1...dx_n~ \Tr_{aux} \Big[B\mathcal{P}\Big\{ (e^{\int _0^L dx Q(x)} \otimes I) (R(x_1)...R(x_n) \otimes O^\dagger(x_1)...O^\dagger(x_n)) \Big\}\Big] \ket{B_{a,\beta}}
        \\
        =& \Tr_{aux}\Big[B\mathcal{P} \Big\{  e^{\int_0^L dx~Q(x) \otimes I } \sum_{n=0}^\infty \frac{1}{n!} \int _0^L dx_1...dx_n~ R(x_1)...R(x_n) \otimes O^\dagger(x_1)...O^\dagger(x_n) \Big\}\Big] \ket{B_{a,\beta}}
        \\
        =& \Tr_{aux}\Big[B\mathcal{P} \Big\{  e^{\int_0^L dx~Q(x) \otimes I } \sum_{n=0}^\infty \frac{1}{n!} \big(\int _0^L dx R(x) \otimes O^\dagger(x) \big)^n\Big\}\Big] \ket{B_{a,\beta}}
    \end{split}
\end{equation}
and finally one can find the representation of the boundary cMPS (BCMPS) as 
\begin{equation}
    \ket{\Psi_{a,\beta}} = \Tr _{aux} \Big[ B \mathcal{P} \exp \int_0^L dx~ \big(Q(x)\otimes I + R(x) \otimes O^\dagger(x)\big)\Big] \ket{B_{a,\beta}}.
\end{equation}

To have an idea for the operator $O(x)$ defined in \eqref{2222}, let us consider the relation 
\begin{equation}
    (T(z) - T(\bar{z})) \ket{B} =0
\end{equation}
for the boundary states. 

By using the relation for the chiral and anti-chiral fields in 2 d
\begin{equation}
    \begin{split}
        2 \pi& T_{zz} (z, \bar{z}) = T(z)
        \\
       2 \pi& \bar{T}_{\bar{z}\bar{z}} (z, \bar{z}) = \bar{T}(\bar{z}) 
    \end{split}
\end{equation}
and expand them in terms of the spacetime components of the energy-momentum tensor
\begin{equation}
    \begin{split}
    T_{zz}= &\frac{1}{4} (T_{00} - 2 i T_{10} - T_{11})
        \\
        T_{\bar{z}\bar{z}}=& \frac{1}{4} (T_{00} - 2 i T_{10} - T_{11})
    \end{split}
\end{equation}
one reaches to the relation 
\begin{equation}
    T_{tx} \ket{B} =0
\end{equation}
Therefore the operator $O$ in the definition of BCMPS is 
\begin{equation}
    O = T_{tx}.
\end{equation}


\section{Holographic interpretation of the boundary cMPS}

\subsection{Continuous measurement}

In this section, based on \cite{osborne2010holographic}, we provide a natural physical interpretation of this variational class.


In the context of cavity-QED, we can directly understand the bulk and boundary fields as follows\cite{schon2005sequential, gardiner2004quantum, kimble1998strong, hijlkema2007single}: think of the cavity modes as the auxiliary system and the quantum field as describing the photons escaping from the cavity.

One can start by describing how we measure something, like a physical observable "M", on a quantum system with $ D$ levels. This approach is called "von Neumann's prescription" \cite{edwards1979mathematical}.
We attach a quantum system with a continuous degree of
freedom, known as meter, in a fiducial state vector $\ket{0}$ and
couple it with the system for some time t according to
the interaction $ H_I = M \otimes p$. If initially, the system is in the state $ \ket{\phi}$, then after the interaction the state is 
\begin{equation}
    e^{-it H_I} \ket{\phi} \ket{0} = \sum_{j=0}^D \phi_j \ket{m_j} \ket{x = m_j t}
\end{equation}
while $ M \ket{m_j }= m_j \ket{m_j}$ and the initial state in the basis of the eigenstate of $M$ is written as 
$ \ket{\phi} = \sum_{j=1} ^D \phi_j \ket{m_j}$.

 The main idea in \cite{osborne2010holographic} is to reverse von Neumann's measurement approach. Instead of focusing on the system as the primary element, we treat the meter as the central system A, and the original system becomes an extra part B. This approach allows us to view it as a state creation tool: we can create various quantum states for meter A by using the measurement approach and then either remove or measure system B. This way, we can generate quantum states for a system with a continuously changing characteristic. 

To proceed, let us consider a family of $D \times D$ complex matrices $R(x),~ x \in [0,L]$ which we measure at time $t=x$ on B. B additionally evolve with a Hamiltonian $K(x)$. The total Hamiltonian is given by 
\begin{equation}\label{ham}
    H(t) = K(t) \otimes I + H_I
\end{equation}
where $ H_I = i R(x) \otimes O^\dagger(x) + h.c.$. Integrating the Schrodinger equation for \eqref{ham}
we get 
\begin{equation}\label{unitary}
    U(L) = \mathcal{P} \exp -i \int_0^L ds \Big( K(s) \otimes I + iR(s) \otimes O^\dagger(s) -i R^\dagger (s) \otimes O(s)\Big)
\end{equation}
The evolution \eqref{unitary}
prepare the class of BCMPS. If we initialize the meter A in the specific boundary state $ \ket{B_{a,\beta}}_A$ and system B in the initial state $ \ket{v_i}$ we have
\begin{equation}
    U(L) \ket{v_i} \otimes \ket{B_{a,\beta}}_A
\end{equation}
Using the Baker-Hausdroff formula, we have 
\begin{equation}
    \begin{split}
        &\exp\Big(ds \big( K(s) \otimes I + R(s) \otimes O^\dagger(s) -i R^\dagger (s) \otimes O(s)\big)\Big) 
        \\
        =& \exp\Big(ds \big( K(s) \otimes I + R(s) \otimes O^\dagger(s)\big)\Big)
        \\
        & \qquad \times \exp\Big(ds -i R^\dagger (s) \otimes O(s)\Big)
        \\
        & \qquad\times \exp \Big( \frac{1}{2} ds ds' ~ [i R^\dagger (s) \otimes O(s),K(s') \otimes I + R(s') \otimes O^\dagger(s') ]\Big)
        \\
        =& \exp\Big(ds \big( K(s) \otimes I + R(s) \otimes O^\dagger(s)\big)\Big)
        \\
        & \qquad \times \exp\Big(ds -i R^\dagger (s) \otimes O(s)\Big)
        \\
        & \qquad\times \exp \Big( -\frac{1}{2} ds R^\dagger(s) R(s) \otimes I + \frac{1}{2} ds ds' [iR^\dagger(s), K(s')] \otimes O(s) 
        \\
        & \qquad\qquad \frac{1}{2} ds ds' [iR^\dagger(s), iR(s')] \otimes O^\dagger (s')O(s)+...\Big).
    \end{split}
\end{equation}
Consider the fact that 
\begin{equation}
    e^{O(x)}\ket{B_{a,\beta}}= \ket{B_{a,\beta}},
\end{equation}
we reach to 
\begin{equation}
    U(L,0) \ket{v_i} \otimes \ket{B_{a,\beta}} =  \mathcal{P} \exp -i \int_0^L ds \Big( Q(s) \otimes I + R(s) \otimes O^\dagger(s)\Big) \ket{v_i} \otimes \ket{B_{a,\beta}}
\end{equation}
while 
\begin{equation}
    Q(x) = -i K(x) - \frac{1}{2} R^\dagger (x) R(x) .
\end{equation}
After projecting the system B on the final state $ \ket{v_f}$, we will reach to the class of BCMPS
\begin{equation}
    \ket{\Psi_{a,\beta}}= \bra{v_f}U(L,0) \ket{v_i} \ket{B_{a,\beta}} = \Tr_{B}\Big[ B\mathcal{P} \exp -i \int_0^L ds \Big( Q(s) \otimes I + R(s) \otimes O^\dagger(s)\Big) \Big] \ket{B_{a,\beta}}
\end{equation}
while the matrix $B$ here is 
\begin{equation}
    B = \ket{v_i}\bra{v_f}.
\end{equation}
Thus, a BCMPS like the class of cMPS can be found from a continuous measurement and the dynamic of system B described by a Lindblad equation.

\subsection{A toy model for evaporating black hole}

In Sec. \ref{3.3} we saw that a given regularized boundary state of a CFT can be written as a TFD state of two CFTs, let us refer to them as left and right CFTs while right CFT stands for the original one, projecting on the corresponding boundary state. 

In Sec. \ref{3.3}, we discussed that at high temperature a regularized boundary state is dual to a microstate of a single-sided black hole. Therefore, the class of BCMPS can be dual to the microstate of the black hole coupled to an ancilla that can represent a bath that absorbs Hawking radiation. 

One can replace the BCMPS as bellow:

\begin{figure}[h]
    \centering
    \includegraphics[width=.6\textwidth]{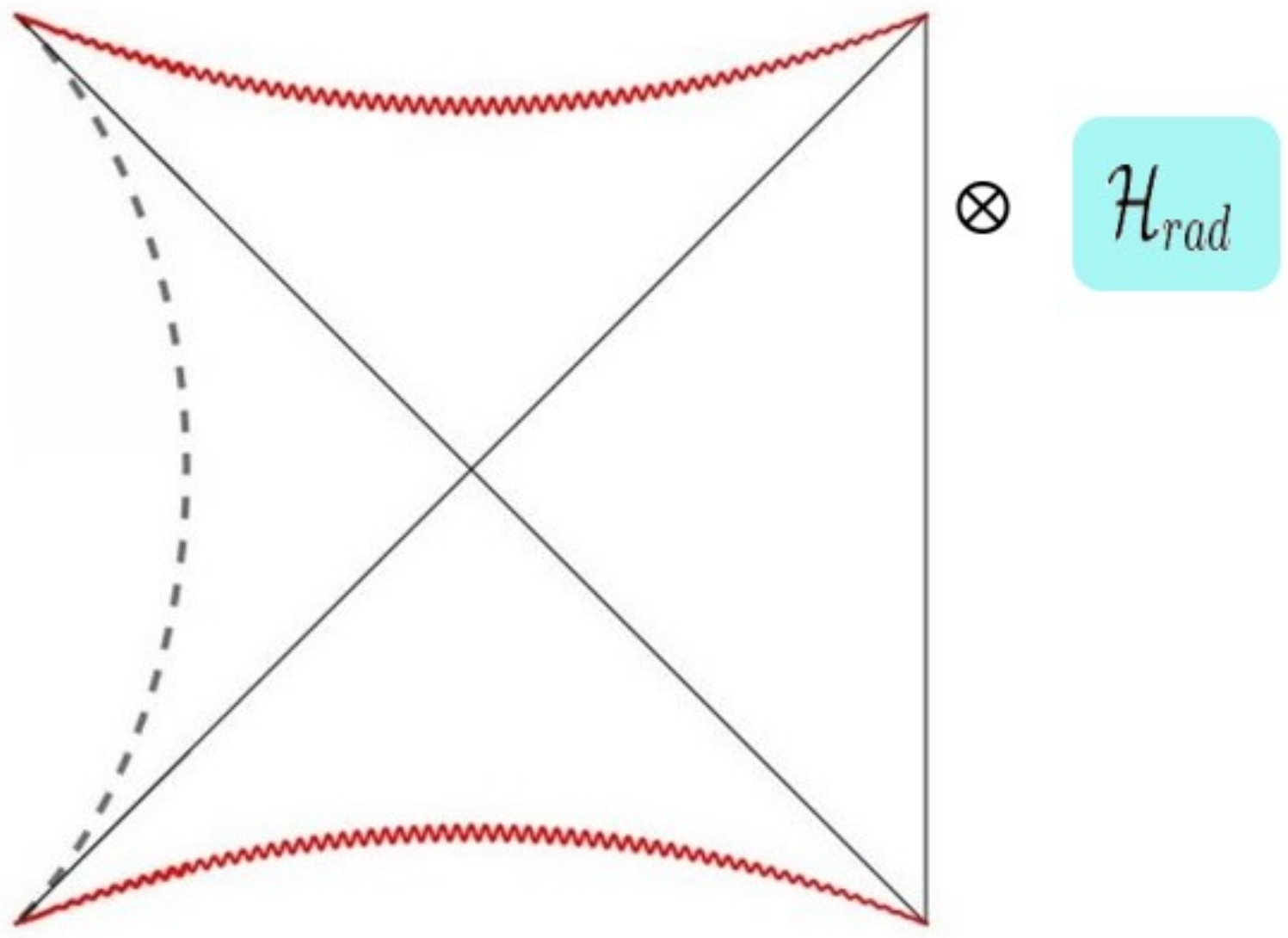}
    \caption{ BCMPS as a toy model for evaporating black hole }
    \label{fig2}
\end{figure}

\begin{equation}
\begin{split}
      \bra{v_f} U(L,0) \ket{v_i} \ket{B_{a,\beta}}
     &=
    \bra{v_f} \bra{B_a} U(L,0)\otimes I \ket{v_i} \ket{TFD(\beta)} 
    \\
   & = \bra{v_f} \bra{B_a} \mathcal{P} \exp \Big( \int_0^L dt~ Q(t) \otimes I _{LR} + R(t) \otimes O ^\dagger_R (t) \otimes I _L\Big)\ket{v_i} \ket{TFD(\beta)}
\end{split}
\end{equation}
while $ Q(t) = -i K(t) -\frac{1}{2} R^\dagger (t) R(t)$.
The interpretation is that at the time  $ t= 0$
we put the ancilla in the state  $ \ket{v_i}$ and two CFTs in the TFD state. They coupled together via the evolution 
\begin{equation}
    H_{tot}= K(t) \otimes I_{LR} + i R(t) \otimes O ^\dagger_R (t) \otimes I _L -  i R^\dagger(t) \otimes O_R (t) \otimes I _L
\end{equation}
the particle created by $O^\dagger$ interchange between the right CFT and ancilla corresponds to the matrix $R$.
Thus, the dual of the BCMPS is a black hole microstate coupled with an ancilla which is a quantum mechanical system.
They both together can be a good system to model the evaporating black holes that are coupled with a bath. The Hawking radiation in such models is absorbed in the bath.
Therefore here the particle goes from the right CFT to the bath or ancilla are holographic dual of the absorption of the Hawking radiation in the bath.

\section{Discussion}

In this paper, we could obtain the new class of BCMPS by repeating the same logic as the original cMPS in the regularized boundary states. It has been discussed that they can have a geometry dual in $2+1$ dimensions as a black hole microstate coupled to an ancilla which plays the role of a system that absorbs the Hawking radiation.
Thus, the BCMPS can be a very good toy model for studying the one-sided evaporating black hole.
Since MPS and its generalization are suitable in $1+1$ dimensions, the BCMPS provides a toy model to study the evaporation of a one-sided black hole just within the AdS$_3$/CFT$_2$. To extend this work to the higher dimension, it might be possible to generalize the class of PEPS. 
By using the class of BCMPS, one can find the $n-$point function so much easier as it inherits the logic of computation of the  $n-$point function from the ordinary cMPS. 

As it has been mentioned, MERA is the most popular class of tensor network which can has geometry dual as AdS. In \cite{miyaji2015boundary},  authors introduce a continuous MERA description of regularized boundary states. 
Having this in mind and replacing the regularized boundary state in the definition of the BCMPS, we expect to have a geometry that looks like the AdS with some modifications that come from the cMPS-like term in the definition of BCMPS.
As we know the black hole microstates near the boundary also have the same geometry as AdS.

On the other hand, the basic principle of quantum error correction (QEC) is to encode information into the long-range correlations of entangled quantum many-body states in such a way that it can not be accessed locally. The relation between  QEC and tensor networks, and particularly MPS has been explored and deepened in some work as \cite{ferris2014tensor}. 
It has been known recently that bulk reconstruction, in particular entanglement wedge reconstruction, is an example of QEC \cite{almheiri2015bulk, Vardian:2023fce, Bahiru:2022ukn}. Moreover, in the case of the evaporating black hole after page time a big portion of the interior called the island is in the entanglement wedge of the ancilla, i.e. the island is encoded in the early Hawking radiation which absorbs in the ancilla. So one can use a technique from QEC called Petz map to reconstruct the operators localized in the island from the ancilla. Having a toy model of the evaporating black hole which is modeled through the BCMPS one can study the reconstruction of the island more concrete.

\acknowledgments
I would like to thank A. Mollabashi and T. Takayanagi for useful discussions and communication. The research is partially supported by the INFN Iniziativa Specifica- String Theory and Fundamental Interactions project.

\bibliographystyle{ieeetr}
\bibliography{refs}

\end{document}